\documentclass[aps,twocolumn,showpacs,preprintnumbers,amsmath,amssymb]{revtex4}
\usepackage{xspace} 

\usepackage{graphicx}
\usepackage{color}
\definecolor{navyblue}{rgb}{0.3,0.3,1}
\definecolor{purple}{rgb}{0.6,0,0.5}
\usepackage[colorlinks=true, pdfstartview=FitV, linkcolor=purple, citecolor=
blue,urlcolor=navyblue]{hyperref}

\newcommand{\be}{\begin{equation}}
\newcommand{\ee}{\end{equation}}

\newcommand{\bee}{\begin{eqnarray}}
\newcommand{\eee}{\end{eqnarray}}

\def\CP{\ensuremath{C\!P}\xspace}
\def\CPT{\ensuremath{C\!PT}\xspace}

\def\be{\begin{eqnarray} &&}

\def\ee{\end{eqnarray}}
\def\bew{\begin{widetext}}
\def\ew{\end{widetext}}

\begin{document}
\title{\CP violation and \CPT invariance in $B^\pm$ decays with final 
state interactions}
\author{I. Bediaga}
\affiliation{Centro Brasileiro de Pesquisas F\'\i sicas, 22290-180, Rio de Janeiro, RJ,
Brazil}
\author{T. Frederico}
\affiliation{Instituto Tecn\'ologico de Aeron\'atica, DCTA,
12228-900, S\~ao Jos\'e dos Campos, SP, Brazil}
\author{O. Louren\c co}
\affiliation{Departamento de Ci\^encias da Natureza, Matem\'atica e Educa\c
c\~ao, CCA, Universidade Federal de S\~ao Carlos,
13600-970, Araras, SP, Brazil}
%\date{\today}
\begin{abstract}
We show that, besides the usual short distance contribution for \CP violation,
final state interactions together with \CPT invariance can play an important role
in  the recent observation of \CP violation in three-body charmless $B^\pm$ decays. A
significant part of the observed \CP asymmetry distribution in the Dalitz plot is located in a
region where hadronic channels are strongly coupled. We illustrate our discussion comparing the
recent observation of \CP violation in the $B^\pm\to K^\pm K^+ K^-$  and $B^\pm\to K^\pm \pi^+
\pi^-$ phase space, with a calculation based on $\pi\pi\to KK$ scattering.
\end{abstract}

\maketitle

\section{Introduction}

For \CP violation to occur, two interfering amplitudes with different weak phases are 
necessary. Until now, all observed \CP violation is compatible with the CKM weak phase,
however there are many modes with interfering amplitudes that produce this asymmetry. For
neutral mesons, direct and indirect \CP asymmetries were observed, the latter associated
to $M^0-\bar M^0$  oscillation, where $M^0 = K^0$  and $B^0$. On the other hand, for 
charged mesons, direct \CP violation was observed only in bottom mesons decays
\cite{babarkpipi,bellekpipi,LHCbD,LHCbConf,LHCbhep,LHCbpipipi}.

The most common mechanism, at the quark level, expected to give a \CP asymmetry in charmless
charged $B$ decays, comes from the short distance BSS model \cite{BSS}, through 
the interference  of the tree
and penguin amplitudes. However, at the hadronic level, there are other interfering
contributions with different weak phases. One of them associated with the interference between
intermediate states, in three-or-more-body decays
\cite{Mirandizing1,Mirandizing2,Gronau2013,Cheng,Chengb}. In general, interference occurs when
two resonant intermediate states, with different weak phases, share the same kinematical region
and hadronic final state. Another possibility is related to hadronic  rescattering in two
different states \cite{wolfenstein,bigi_book}.

Wolfenstein \cite{wolfenstein}, based on \CPT
invariance and unitarity, proposed a formalism for decay, in which the hadronic final-state
interaction (FSI) and \CPT constraint are considered together. From that, the sum of the
partial widths for channels coupled by the strong Hamiltonian, must be equal to the
corresponding sum of the partial decay widths of the associated  anti-particle. It is  more
restrictive than the  \CPT condition, which equates the lifetime for a particle and its
anti-particle.  Then, in addition to the usual
\CP-violating amplitude from the BSS mechanism, one has the asymmetry induced by rescattering,
namely the ``compound" contribution~\cite{Soni2005}.

The large number of  final states with the same flavour quantum numbers, accessible for a
charmless $B$ meson decay, could wash out the ``compound" contribution for a single decay
channel. However, since hadronic many-body rescattering effects are far from being understood,
it is evident that this phenomenological hypothesis deserves to be tested experimentally and
further explored theoretically. The aim of this paper is to investigate the possible presence
of the ``compound" contribution in charmless three-body charged $B$ decays presented recently
by the LHCb collaboration~\cite{LHCbConf,LHCbhep}.

\section{Basic facts and our assumptions} 

One of the most intriguing characteristics in three-body charmless $B$ decays, observed by
Belle~\cite{bellekpipi}, Babar~\cite{babarkpipi} and now by LHCb~\cite{LHCbhep}, is that
the two-body distributions of events are concentrated at low invariant mass taking into
account the huge phase-space available, for example, in $B^{\pm}\to\, K^{\pm}\pi^+\pi^-$.
 The distributions of events in $K^{\pm}\pi^{\mp}$ and $\pi^+\pi^-$ invariant masses
squared are mostly concentrated below 3~GeV$^2$ (except charmonium intermediate states).
This result confirms the old phenomenological assumption of the isobar model, in which the
final state factorizes in a two body interacting system plus a bachelor. In this case, 
the rescattering associated to  hadron-hadron interactions should be below the experimental
limit of 3~GeV$^2$, that is basically in the elastic hadron-hadron regime \cite{Gronau2013}. 

The two-body elastic scattering data from different collaborations in the 70's and 80's 
can be well parametrized within S-matrix theory.  
The opening of new channels is encoded by the inelasticity $(\eta)$, which
represents  the amount of two-body elastic flux lost at a given energy. For
$\eta=1$ no inelastic processes happen. In general, the S-matrix element is
represented by the unitary Argand diagram, which allows to
identify resonances through  phase variation and also the inelasticity. 
If data are around a circle, $\eta=1$, otherwise appear inside the circle and 
inelastic scattering takes place.

The Argand plot for S-wave $\pi \pi$ elastic scattering from the CERN-Munich
collaboration \cite{CERN-Munich} shows $\eta$ close to one up to
$f_0(980)$, after that $\eta < 1$ and then
returns to one for masses above  1.4 GeV. The deviation from the unitary circle at
1 GeV is explained by  $\pi\pi$ coupling to $KK$ channels. Experimental
results from the early 80's show an important S-wave $\pi\pi\to KK$ scattering between 1
and 1.6 GeV \cite{Cohen1980}, with a corresponding decrease of the S-wave
$\pi\pi$ elastic amplitude \cite{Surovtsev2010}.
The observed inelasticity of the $\pi\pi$ S-wave amplitude is basically associated
only to the $\pi\pi\to KK$  process (see also the analysis presented in
Ref.~\cite{Pelaez06}).
 For the P-wave, the CERN-Munich
experimental results show $\eta=1$ until 1.4 GeV. Then $\eta$
drops to a minimum of 0.5, due to the presence of the $\rho(1690)$, which prefers to decay
into four pions. Finally, the  D-wave is elastic
until 1.2 GeV, after that $\eta$ slowly decreases. In short, the $\pi\pi\to\pi\pi$ scattering, except 
for the S-wave in the invariant mass region of $1\lesssim m_{\pi\pi}\lesssim 1.5$  GeV, the elastic
scattering is the dominant contribution.

The other important study is the $K\pi\to K\pi$ scattering from the LASS
experiment~\cite{LASS}. The S-wave has inelastic events above $1.5$~GeV, and it has both
isospin $1/2$ and $3/2$ states. The P-wave is elastic up to $1.41$~GeV and inelastic when
$K^*(1680)$ is formed, as it  can decay to $K\rho$ and $K^*\pi$. Finally, the D-wave is
elastic in a small region and is dominated by $K_2(1430)$, which decays to $K\pi$ about
half of the time.

The conjunction between: i) the general hypothesis of dominant 2 + 1 processes in
charmless three-body $B$ decays, supported by the observed distribution of the Dalitz
plot, basically, at very low hadron-hadron masses; and ii) the observed dominance of the
hadron-hadron elastic scattering, in the same region
where the majority of the two-body decays are placed in the Dalitz plot, allows us to
assume that the rescattering effects in three-body $B$ decays happen essentially in
$3\to 3$ channels. Some small contributions from D-wave can also be added to $3\to5$
process, but for our general purpose it can be neglected. More sophisticated processes
such as the rescattering involving the bachelor particle can be added, but they must be
understood as a correction to the main contribution coming from  2+1 processes
\cite{Bras2011}.

Note that this conclusion can be used only for three-body decays,
because we know well the events distribution in the Dalitz plot. The same argument does
not fit for two-body charmless $B$ decays. In that case one has to understand what is the
contribution to the hadron-hadron elastic scattering in the $B$ mass region, which 
 is not yet available experimentally. Also for four-body decays, we do not have a
clear experimental picture for two or three-body mass distributions.

Our working assumption, based on experimental evidences from $\pi\pi$ and $KK$ scattering,
is to investigate the effect of two-body rescattering contributions to the 
\CP-violating charged
$B$ decays in the strongly coupled $\pi\pi$ and $KK$ channels. 

\section{\CPT invariance in a  decay} 

To define our notation and the framework for implementing the \CPT constraint in  $B$
meson decays, we follow closely Ref.~\cite{Marshak,Branco}. A hadron
state $|h\rangle$ transforms under \CPT as ${\mathcal {CPT} }\,|h\rangle=\chi\langle
\overline h|$, where $\overline h$  is the charge conjugate state and $\chi$ a phase.
The weak and strong Hamiltonians conserve \CPT, therefore
%\begin{eqnarray}
$({\mathcal {CPT}})^{-1}\,H_w\,{\mathcal {CPT} }= H_w $ and  %\,\,\,\mbox{and}\,\,\, 
$({\mathcal {CPT} })^{-1}\,H_s\,{\mathcal {CPT} }= H_s. $%\,\,
%\label{cpt2}
%\end{eqnarray}
The weak matrix element for the hadron decay is
$\langle \lambda_{out}|H_w|h\rangle$,
where $\lambda_{out}$ includes the distortion from the strong force due to the final state
interaction. The requirement of \CPT invariance is fulfilled for the matrix element when
\begin{eqnarray}
\langle \lambda_{out}|H_w|h\rangle
= \chi_h\chi_\lambda\langle  \overline \lambda_{in} |H_w| \overline h  \rangle^*\,.
\label{cpt3}
\end{eqnarray} 
Inserting the completeness of the strongly interacting states, eigenstates of $H_s$, and
using hermiticity of $H_w$, one gets
\begin{eqnarray}
\langle \lambda_{out}|H_w|h\rangle
=\chi_h\chi_\lambda\sum_{\overline\lambda^\prime} S_{\overline\lambda^\prime,\overline\lambda}
\langle \overline\lambda^{\prime}_{out} |H_w| \overline h  \rangle^* ,
\label{cpt4}
\end{eqnarray} 
where the S-matrix element is $S_{\overline\lambda^\prime,\overline\lambda}=
\langle \overline \lambda^\prime_{out}|\overline \lambda_{in}\rangle$.
 
The sum of partial decays width of the hadron decay and the correspondent sum for the charge conjugate  should
be identical, which follows from Eq.~(\ref{cpt4})
% \begin{small}
\begin{eqnarray}
\sum_\lambda|\langle \lambda_{out}|H_w|h\rangle|^2&=&\sum_{\overline\lambda}
|\sum_{\overline\lambda^\prime}S^*_{\overline\lambda^\prime,\overline\lambda}
\langle \overline\lambda^{\prime}_{out} |H_w| \overline h  \rangle |^2 \nonumber\\
&=&\sum_{\overline\lambda } 
|\langle \overline\lambda_{out} |H_w| \overline h \rangle|^2,\,\,\,\,\,\,\,\,
\label{cpt5}
\end{eqnarray} 
% \end{small}
and note that besides the \CPT constraint we have also used the hermiticity of the weak
Hamiltonian.

The \CP-violating phase enters linearly at
lowest order in the hadron decay amplitude. In general, the decay amplitude can be written
 as $\mathcal{A}^{\pm} = A_\lambda + B_\lambda e^{\pm i\gamma}$, where
$A_\lambda$ and $B_\lambda$ are complex amplitudes invariant under \CP, containing the
strongly interacting final-state channel, i.e., $\mathcal{A}^{-}=\langle\lambda_{out}
|H_w| h  \rangle$, and $\mathcal{A}^{+}=\langle \overline \lambda_{out} |H_w|  \overline h
 \rangle$. The only change due to the \CP transformation is the sign
multiplying the weak phase $\gamma$. 

\section{Coupled-channel  decay, \CPT and \CP asymmetry} 

Now, we discuss the example of a decay to channels coupled by rescattering, i.e., the
strong S-matrix has non-vanishing off-diagonal matrix elements,
$S_{\lambda^\prime,\lambda}=\delta_{\lambda^\prime,\lambda}+i\,
t_{\lambda^\prime,\lambda}$, where $t_{\lambda^\prime,\lambda} $ is the strong scattering
amplitude of $\lambda^\prime\to\lambda$, and $\delta_{\lambda^\prime,\lambda}$ is the
Kronecker delta symbol. In this case the \CPT condition (\ref{cpt5}) gives
\begin{eqnarray}
\sum_\lambda \Gamma(A^-_\lambda)  = \sum_{\overline \lambda} \Gamma(A^+_{\overline \lambda} ) \ , 
\label{cp5} 
\end{eqnarray}
where the subindex labels the final state channels, summed up in the kinematically 
allowed phase-space.

The decay amplitude  written in terms of the \CPT constraint (\ref{cpt4}), and considering the
\CP violating amplitudes for the hadron and its charge conjugate, is given by
\begin{eqnarray}
A_\lambda+e^{\mp i \,\gamma}B_\lambda
=\chi_h\chi_\lambda\sum_{\lambda^\prime} S_{\lambda^\prime,\lambda}
\left(A _{\lambda^\prime}+e^{\pm i\,\gamma}B_{\lambda^\prime}\right)^*
\, .
\label{cp11}\end{eqnarray} 
Note that the above equation imposes a relation between $A_\lambda$ or $B_\lambda$
with their respective complex conjugates.

\section{\CP asymmetry and FSI at leading order} 

The full decay amplitudes $A_\lambda$ and $B_\lambda$ can be separated in two parts, one
carrying the FSI distortion ($\delta A_\lambda$, $\delta B_\lambda$) and another
one corresponding to a source term without FSI ($A_{0\lambda}$, $B_{0\lambda}$), $
A_\lambda = A_{0\lambda} + \delta A_\lambda$ and $B_\lambda =B_{0\lambda} + \delta
B_\lambda$. Retaining terms up to leading order (LO) in $t_{\lambda^\prime,\lambda}$ in
(\ref{cp11}), one can easily find that
\begin{eqnarray}
\mathcal{A}^+_{LO}=A_{0\lambda}+e^{i\gamma}B_{0\lambda} +
i\sum_{\lambda^\prime}t_{\lambda^\prime,\lambda}
\left(A _{0\lambda^\prime}
+e^{i\gamma}B_{0\lambda^\prime}\right),
\label{cp24}
\end{eqnarray} 
where we have used that 
\begin{eqnarray}
A_{0\lambda}&=&\chi_h\chi_\lambda A_{0\lambda}^*\quad \mbox{and} 
\label{a0b01} \\
B_{0\lambda}&=&\chi_h\chi_\lambda B_{0\lambda}^*, 
\label{a0b02} 
\end{eqnarray} 
which come from
(\ref{cpt3}), when the strong interaction is turned off. We point out that
Eq.~(\ref{cp24}) is equivalent to the shown in~\cite{wolfenstein,bigi_book}, but it was obtained
with a different approach.  
 
The \CP asymmetry, $\Delta \Gamma_\lambda=\Gamma\left(h\to \lambda\right)-\Gamma(\overline
h\to \overline\lambda)$, evaluated by considering the amplitude (\ref{cp24}) and only
terms up to leading order in $t_{\lambda^\prime,\lambda}$, is given by
% \begin{small}
\begin{eqnarray}
\Delta \Gamma_\lambda
&=&4(\sin\gamma) \, \mbox{Im}[ B^*_{0\lambda}A_{0\lambda} \nonumber\\
&+&i \sum_{\lambda^\prime}(B^*_{0\lambda}t_{\lambda^\prime,\lambda}A
_{0\lambda^\prime} - B^*_{0\lambda^\prime}t^*_{\lambda^\prime,\lambda}A
_{0\lambda})],
\label{cp26}
\end{eqnarray}
% \end{small}
where the external sum of $\lambda^\prime$ represents each  channel separately. The second
and third terms in the imaginary part in Eq.~(\ref{cp26}) can be associated to the
``compound " \CP asymmetry~\cite{Soni2005}, and have the important property of canceling
each other when summed with all FSI, in order to satisfies the \CPT condition
expressed by Eq.~(\ref{cp5}). The first term, namely $B^*_{0\lambda}A_{0\lambda} $, is
related to the interference between  two \CP conserving amplitudes without FSI, as happens
for the tree and penguin amplitudes in the BSS model \cite{BSS}. This term must satisfy
\begin{eqnarray}
\sum_\lambda \mbox{Im} \left[B_{0\lambda} \,A^*_{0\lambda} \right] = 0,
\label{cp22}
\end{eqnarray}
as a consequence of the \CPT constraint. 

The cancellation in Eq.~(\ref{cp22}) reflects the stringent condition of \CPT invariance
given in Eq.~(\ref{cpt3}), when the FSI is turned off.  Therefore, the general condition
given by Eq.~(\ref{cp22}) should be satisfied, with  one  trivial solution that  the phase
difference between the two \CP-conserving  amplitudes is zero for all decay channels. This
term was neglected by Wolfenstein. 

Noteworthy to mention here that the second term in Eq.~(\ref{cp26}) also satisfies the
\CPT condition, which follows straightforwardly by using Eqs.~(\ref{a0b01})-(\ref{a0b02}),
the symmetry of $t_{\lambda,\lambda^\prime}$, and the fact that the strong interaction
does not mix different $CP$ eigenstates.

\section{Inelasticity and \CP violation in a two-channel problem.} 

Considering the case of two body and two coupled channels, $\alpha$ and $\beta$, the
unitarity of the S-matrix together with its symmetry ($S_{\alpha,\beta}=S_{
\beta,\alpha}$), leads to $|S_{\alpha\alpha}|^2+ |t_{\beta,\alpha}|^2=|S_{\beta\beta}|^2+
|t_{\beta, \alpha}|^2=1$ and
$S_{\alpha\alpha}\,t^*_{\beta,\alpha}-S_{\beta\beta}^*\,t_{\beta,\alpha}=0$.
By writing the diagonal elements of the two body elastic scattering S-matrix as 
$S_{\alpha\alpha}=\eta_\alpha e^{2i\delta_\alpha}$ and $S_{\beta\beta}=\eta_\beta
e^{2i\delta_\beta}$, where $\eta_\alpha$ and $\eta_\beta$ are the inelasticity 
for the $\alpha$ and $\beta$ channels, respectively, one gets that
$\eta_\alpha=\eta_\beta=\eta$, and $|t_{\beta,\alpha}|=\sqrt{1-\eta^2}$.
Furthermore, one can easily derive that $t_{\beta
,\alpha}=\sqrt{1-\eta^2}\,e^{i\left(\delta_\alpha+\delta_\beta\right)}$. Therefore, we can
rewrite Eq.~(\ref{cp26}) for the $\alpha$ channel as a sum of two distinct
terms, namely, the short distance and the compound contributions. The expression can be written
as
\begin{equation}
\Delta \Gamma_\alpha
=4(\sin\gamma)\left( \zeta_0 + \sqrt{1-\eta^2}\, \zeta_1\right) .
\label{cp32}
\end{equation}
The term containing  
\begin{eqnarray}
\zeta_0=\mbox{Im}
\left[B^*_{0\alpha}A_{0\alpha}(1+i(t_{\alpha\alpha}-t^*_{
\alpha\alpha}))\right] \ 
\label{zeta0}
\end{eqnarray}
corresponds to the short distance
contribution to the \CP asymmetry. It is widely used to calculate \CP asymmetries in two-body
$B$ decays, through the interference between the tree and penguin amplitudes for single
decays. 

The term corresponding to the compound contribution in Eq.~(\ref{cp32}) contains
\begin{eqnarray}
\zeta_{1}=|K_\alpha|\cos\left(\delta_\alpha+\delta_\beta + \Phi_\alpha \right),
\label{cp33}
\end{eqnarray}
where $K_\alpha=B^*_{0\alpha}\, A_{0\beta}- B _{0\beta}\,A^* _{0\alpha}$ and $\Phi_\alpha=-i\ln
(K_\alpha/|K_\alpha|)$. This non diagonal term  gives a close relation between the region for
\CP-violation and inelastic $\alpha\to\beta$ scattering, presented above. We remind that the
opposite sign of $\Delta \Gamma_\beta$ in respect to $\Delta\Gamma_\alpha$ comes from
Eqs.~(\ref{a0b01})-(\ref{a0b02}), and that the strong interaction does not mix states with
different phases $\chi_\lambda$, which leads to 
\begin{eqnarray}
K_\beta=-K_\alpha \quad \mbox{and}\quad \Phi_\beta=\Phi_\alpha+\pi.
\label{phialpha}
\end{eqnarray}
Note that from Eq. (\ref{cp22}) applied to the two-channel case, the short 
distance term
satisfies $\Delta\Gamma_\alpha=-\Delta\Gamma_\beta$,  which is also verified for the compound
contribution as a consequence of Eq.~(\ref{phialpha}), discussed above.

Indeed, looking at the LHCb results \cite{LHCbConf,LHCbhep},  a direct and complementary
relation between different charmless three-body  decay channels coupled by the strong
interaction emerges for $B^\pm\to K^\pm\pi^+ \pi^-$ and $B^\pm\to K^\pm K^+ K^-$, and for the
decays $B^\pm\to \pi^\pm\pi^+ \pi^-$ and $B^\pm\to \pi^\pm K^+ K^-$. Even tough  the tree and
penguin composition in the total decay  amplitudes for each pair of coupled channels are
expected to be different. The \CP asymmetry distribution in the Dalitz plot for these
channels shows the prevalence of  \CP violation in the mass region where the $\pi\pi\to KK$
scattering is important. As a matter of fact, the $\pi^+\pi^-$ and $K^+K^-$ channels are
coupled to $\pi^0\pi^0$ and $K\overline K$. Besides that, the two channels with two or more
kaons in the final state have \CP asymmetries with opposite signs with respect to the ones with
two or more pions.  {These facts motivates us to look more closely to the compound contribution
to the partial decay widths in the three-body $B$ decays.}

\section{Estimate of the compound contribution to $\Delta\Gamma_{KK(\pi\pi)}$ in $B^\pm\to
K^\pm K^+ K^-$ ($K^\pm \pi^+ \pi^-$) decays} 

To perform a simple test of the compound contribution (second term of 
Eq.~(\ref{cp32}))
to \CP asymmetry using only a single angular momentum channel, namely, the S-wave, 
the best
place is to look to the asymmetry in decays involving  $KK$ and $\pi\pi$ channels. Beyond the
$\phi$ mass region, there are no other significant resonance contributions with a strong $KK$
coupling before the $ f_2(1525)$ resonance. Therefore as an illustration, we estimate the
compound contribution to the asymmetry $\Delta\Gamma_{KK(\pi\pi)}$ in $B^\pm\to K ^\pm K^+ K^-$
($K^\pm \pi^+\pi^-$) decays, presented by the LHCb collaboration~\cite{LHCbhep}.

As a remark, the three-body rescattering effect at the two-loop level is small 
compared
to the first two-body collision contribution, as suggested by the
three-body model calculation for the $D^\pm\to K^\pm \pi^+\pi^- $
decays~\cite{Bras2011}. We assume that this approximation for charmless three-body $B$ decays
 must be valid at least for some regions of the phase space.

In order to get a quantitative insight on the enhancement of the \CP asymmetry 
from the
coupling between the $\pi\pi$ and $KK$ channels in the compound contribution, we start by
defining the channels $\alpha\equiv K^+K^-$ and $\beta\equiv\pi^+\pi^-$ and consider the main
isospin channel $I=0$ and $J^P=0^+$. From the second term of  Eq.~(\ref{cp32}) with $\zeta_1$
from Eq.~(\ref{cp33}), we can write the compound contribution to the \CP asymmetry as
\begin{align}
\Delta\Gamma^{\rm comp}_{KK} \approx \mathcal{C}\sqrt{1-\eta^2}\,
\cos\left(\delta_{KK}+\delta_{\pi\pi}+ \Phi_{KK}\right)F(M^2_{KK}),
\label{cp35}
\end{align}
with $\mathcal{C}=4|K|\,(\sin\gamma)$ considered energy independent. We still approximate the
kaon-kaon S-wave phase shift as $\delta_{KK}\approx\delta_{\pi\pi}$ in the region where the
channels are strongly coupled. The Dalitz phase-space factor is
$F(M^2_{KK})=(M^2_{K^+K^-})_{max}-(M^2_{K^+K^-})_{min}$, for the \mbox{$B^\pm\to K^\pm K^+
K^-$} channel (see e.g. \cite{pdg}). The masses $(M^2_{K^+K^-})_{max}$ and
$(M^2_{K^+K^-})_{max}$ depend on the $KK$ subsystem mass, $M^2_{KK}$. Also the
symmetrization of the decay amplitude in the two equally charged kaons is disregarded as
the low mass regions for each possible neutral $KK$ subsystem are widely separated in
phase space. 

Following
Ref.~\cite{pelaprd05}, we have used the parametrization for the pion-pion 
inelasticity and phase-shift, for the $I=0$ and $J^p=0^+$ dominant channel, in order to
evaluate Eq.~(\ref{cp35}). The used parametrizations are given in Ref.~\cite{pelaprd05} by
Eqs.~(2.15a), (2.15b), (2.15b'), (2.16), and the quoted errors.  We also
use the $CPT$ condition given by Eq.~(\ref{cp5}), restricted to two channels, to obtain the
asymmetry in the $\pi\pi$ decay channel, which in this case is given by
$\Delta\Gamma^{\rm comp}_{\pi\pi}=-\Delta\Gamma^{\rm comp}_{KK}$. 

\begin{figure}[!htb]
\includegraphics[scale=0.35]{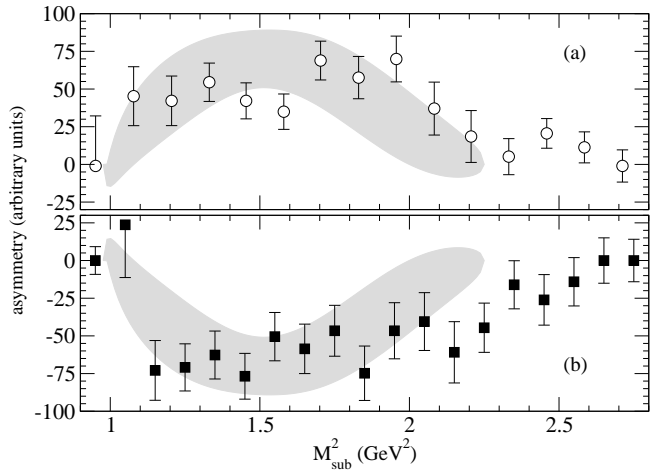}
\caption{Estimate (grey band) of Eq.~(\ref{cp35}) as a function of the
subsystem mass compared to experimental data of (a) the asymmetry of
$B^\pm\to K^\pm\pi^+\pi^-$ decay (circles), and of (b) the asymmetry of $B^\pm\to K^\pm
K^+ K^-$ decay (squares). Data extracted from Ref.~\cite{LHCbhep}.}
\label{fig1}
\end{figure}

In order to compare the asymmetries $\Delta\Gamma^{\rm comp}_{KK}$ and $\Delta\Gamma^{\rm
comp}_{\pi\pi}$ to experimental data, we extracted the difference $B^- - B^+$, respectively for
the $B^\pm\to K^\pm K^+ K^-$  and $B^\pm\to K^\pm \pi^+ \pi^-$ decays, from the recent LHCb
results presented in Ref.~\cite{LHCbhep}. The results are shown in Fig.~\ref{fig1} for an
arbitrary normalization fitted to $\Delta\Gamma^{\rm comp}_{KK}$. Our calculations are
presented from the subsystem mass ($M^2_{\mbox{\tiny sub}}$) above the $KK$ mass threshold.
Indeed, $M^2_{\mbox{\tiny sub}}=M^2_{K^+K⁻}$ ($M^2_{\pi^+\pi⁻}$) for $B^\pm\to K^\pm K^+ K^-$
($K^\pm \pi^+ \pi^-$).

The width of the band represents the errors in the
parametrizations of the isoscalar S-wave $\pi\pi$ phase shift, and inelasticity parameter,
both taken from Ref.~\cite{pelaprd05}. The phase $\Phi_{KK}$ was chosen to be
zero, which emphasizes the role of the strong phases in \CP violation process. Note that
this assumption is accompanied by $\Phi_{\pi\pi}=\pi$ according to the relation given in
Eq.~(\ref{phialpha}), therefore, it is ensured that
$\Delta\Gamma^{\rm comp}_{KK}=-\Delta\Gamma^{\rm comp}_{\pi\pi}$.

We can see  a qualitative agreement between the model parameterized with the $\pi\pi$
elastic phase-shift with data, mainly in the sense that the \CP violation distribution
observed in both $B^\pm\to K^\pm K^+ K^-$ and $B^\pm\to K^\pm\pi^+\pi^-$ decays are
important to the mass region where the S-wave scattering $\pi^+ \pi^- \to K^+ K^-$ is
important, as shown in Fig.~\ref{fig1}. A visual inspection of the Dalitz plot of the
 $B^\pm\to K^+K^-\pi^\pm$ and $B^\pm\to\pi^\pm\pi^+\pi^-$ decays~\cite{LHCbpipipi}, also
presents an important  \CP violation distribution at similar  masses to those where \CP
violation is relevant for $B^\pm\to  K^\pm K^+ K^-$ and $B^\pm\to  K^\pm \pi^+ \pi^-$.
Also the \CP asymmetry below the $KK$ threshold in the resonance region appears
appreciable, which is however outside the region where the FSI mechanism discussed here
applies.

\section{Comments}

Although we have focused only on the relevance of the coupling between $\pi\pi$ and $KK$
channels in the \CP asymmetry observables using the $CPT$ constraint, one should note that 
three light-pseudoscalar mesons can, in principle, couple via strong interaction with channels 
like $D\overline D h$, where $h$ can be $\pi$ or $K$. It seems reasonable to expect 
that $D\overline Dh \to hhh $ can contribute to the \CP asymmetry in regions of large two-body 
invariant mass above the $ D\overline D $ threshold, %assuming the rescattering $ hh \to D\overline D $; 
that is far from the $KK$ threshold and above 1.6 GeV, outside the region 
discussed in this work.
Furthermore, there is no available  experimental data and even theoretical
predictions for these possible long-range interactions to induce \CP asymmetries above $
D\overline D $ threshold in charmless three-body charged decays, as we did using the $\pi\pi
\to KK$ scattering. Since that  direct \CP violation induced by the short distance
interaction must be highly suppressed in double charged charm $B$ decays, future experimental 
analysis could look for those asymmetries in order to observe \CP violation induced by
rescattering originated by charmless $B$ decay channels.

The difficulty to observe this ``compound'' \CP asymmetry in double charm charged $B$ decays
comes because the branching fractions of these decays are  about two orders of magnitude
larger than the corresponding one for charmless $B$ decays. Therefore, in order to measure the
induced \CP asymmetries in double charm charged $B$ decay channels, the \CP violation must be
large enough to overcome the increase in the branching fraction ratios when compared to three
light-pseudoscalar  channels. Despite  the global suppression due the large difference in
branching fractions pointed above, double charm charged three-body $B$ decays, can present a
specific and concentrated phase-space region where the ``compound'' \CP asymmetry takes place.

Although we have compared the data for the asymmetry only to the compound 
contribution,
one must be aware of the first term in Eq. (\ref{cp32}) containing $\zeta_0$, that carries
the short range physics. 
The comparison with the data suggests the importance of the rescattering, which 
seems 
to be relevant in the region of masses analyzed in Fig.~\ref{fig1}. However, the LHCb results 
for charged $K\pi\pi$ and $\pi\pi\pi$ presents a clear \CP violation below the $KK$ threshold, 
and in this region it may be possible to have a more clean access to \CP violation from 
short distance contributions. 

\section{Conclusions} 

We studied \CP violation in three-body charmless $B^\pm$ decays using two
basic assumptions: i) \CPT invariance; and ii) that part of this \CP violation
is due to the interference of two \CP-conserving hadronic amplitudes separated by a
\CP-noninvariant phase. We have built a plausible scenario where these two assumptions lead to
the observed asymmetries in both  $B^\pm\to K^\pm K^+ K^-$ and $B^\pm\to K^\pm \pi^+ \pi^-$
decays as found by the LHCb collaboration~\cite{LHCbhep}, which are also concentrated in the
low $K^+ K^-$ and $\pi^+\pi^-$ mass regions. The coupling between the $KK$ and $\pi\pi$
channels is strong in the energy range where the asymmetry in $B^\pm\to K^\pm K^+ K^-\, ( K^\pm
\pi^+ \pi^-)$ decay is observed, indicating that the ``compound " contribution 
should be
taken into account to reproduce the experimental data. Modulated by a phase-space 
factor, the
asymmetry is proportional to
$\sqrt{1-\eta^2}\cos\left(\delta_{KK}+\delta_{\pi\pi}+\Phi\right)$, coming from the
magnitude and phase of the  $\pi\pi\to KK$ transition amplitude. In the future, the analysis of 
the \CP asymmetry in charmless $B$ decays can be extended to include corrections 
(expected to be small) induced by the three-body rescattering processes.

\section*{Acknowledgments} 

We gratefully acknowledge  I. Bigi, A. Gomes, M. Gronau, 
J. Miranda, I. Nasteva and  A. dos Reis  for
stimulating discussions and valuable comments. We thank the partial support from 
Funda\c c\~ao de Amparo \`a 
Pesquisa do Estado de S\~ao Paulo (FAPESP) and to Conselho Nacional de 
Desenvolvimento Cient\'\i fico e Tecnol\'ogico (CNPq) of Brazil.

\end{document}